\documentstyle[aps,prl,twocolumn,floats,epsf]{revtex}

\newcommand{\eqref}[1]{Eq.~(\protect\ref{#1})}
\newcommand{\figref}[1]{Fig.~\protect\ref{#1}}

\begin{document}

\draft

\wideabs{

\title{
Transition from Spatiotemporal Chaos to Ideal Straight  Rolls\\
in Rayleigh-B\'{e}nard Convection
}

\author{ 
Reha~V. Cakmur\cite{C-address},
David~A. Egolf\cite{CTC-address}, 
Brendan~B. Plapp, and Eberhard Bodenschatz\cite{eb-email},
}

\address{
Laboratory of Atomic and Solid State Physics, Cornell University,
Ithaca, NY 14853-2501
}

\date{\today}

\maketitle

\begin{abstract}

For Rayleigh-B\'{e}nard convection in a square cell with a fluid
of Prandtl number one, we report experimental results on the
transition between a stationary pattern of ideal straight
rolls (ISR) and the spatiotemporal chaotic state of
spiral defect chaos (SDC).  In contrast to experiments in circular
geometries, we found ISR states below a particular value of the control
parameter and SDC states above this value.   By characterizing the
pattern with a global measure, the pattern entropy, we found
that the transition from SDC to ISR showed similarities to phase transitions
in equilibrium finite-size systems.

\end{abstract}

\pacs{
47.54.+r,
47.20.Lz,
47.27.Te
}
}

\narrowtext

In recent years scientists in a number of fields have shown great interest
in dissipative pattern-forming systems of large spatial extent.
These systems often display particularly intriguing states of
persistent time-dependent behavior termed spatiotemporal chaos
\cite{CrossHohenberg93,Gollub94,CrossHohenberg94}.
One important question regarding these states is whether ideas from
statistical mechanics may be useful for understanding the complex
spatiotemporal behavior.  In this Letter we present experimental
data showing that a transition between a spatiotemporal chaotic state
and a well-ordered time-independent state of Rayleigh-B\'{e}nard
convection is similar to phase transitions in finite-size equilibrium
systems \cite{Newman93}.

\begin{figure}   
\centerline{\epsfxsize=3.0in \epsfbox{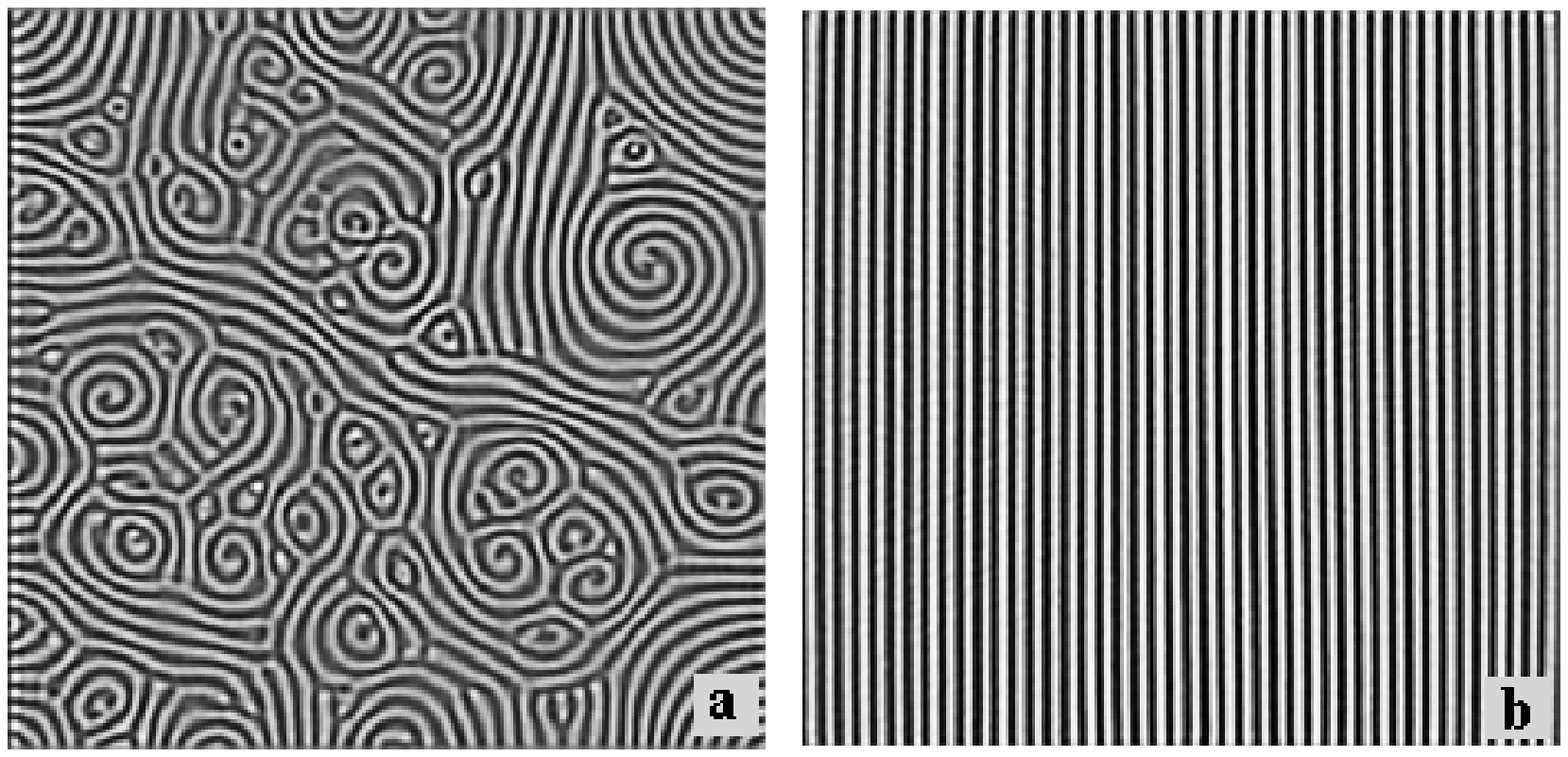}}
\bigskip
\caption{
{\bf (a)} Spiral defect chaos and 
{\bf (b)} ideal rolls at $\epsilon = 0.92$. Warm upflow 
corresponds to dark and cold downflow to bright. 
}
\label{bistable-fig}
\end{figure}

In a Rayleigh-B\'{e}nard convection experiment, a horizontal fluid
layer of height~$d$ is confined between a top plate of
uniform temperature~$T_{\rm top}$ and a bottom plate of
uniform temperature~$T_{\rm bottom} = T_{\rm top} + \Delta T$.
For $\Delta T$ less than a critical value~$\Delta T_c$, the fluid is
stationary; however, when $\Delta T > \Delta T_c$, the stationary state
is unstable and a pattern of convection rolls with wavenumber
$k \approx \pi/d$ develops \cite{Landau59}.  
For systems of infinite extent in the
horizontal directions it was shown theoretically  that
an  ideal pattern of straight rolls (ISR) of wavenumber $k$ is stable in
a restricted region of the $\epsilon-k$ parameter space
\cite{Clever74}, where
$\epsilon = (\Delta T / \Delta T_c - 1)$ is the reduced control parameter.
Experiments with argon gas in a rectangular cell of small 
aspect ratio $\Gamma = ({\rm length}/2) / d \approx 15$
\cite{Croquette89}
showed reasonable agreement with these predictions.  However,
for a larger circular cell of radial aspect ratio $\Gamma = r/d = 76$, 
Morris {\it et al.} found a state of spatiotemporal chaos within the
boundaries of the stable region of the $\epsilon-k$ parameter space
\cite{Morris93}.
This state, termed spiral defect chaos (SDC) and shown in
\figref{bistable-fig}(a), is characterized by a
complex dynamics involving targets, rotating spirals, 
dislocations, disclinations,
and grain boundaries. However, as shown in \figref{bistable-fig}(a),  
SDC  also  shows narrow regions  of slightly curved ISR. 
In this Letter we argue that it is the growth of these regions 
that characterizes the transition from  SDC to ISR.

Since the discovery of SDC an unresolved
question was whether the ISR predicted to
be stable for an infinite system could be realized
in a large aspect ratio experiment.  Recently, Cakmur {\it et al.} used  
specially prepared systems and showed that ISR were 
indeed stable where predicted, and that both SDC states and 
ISR states existed for the same value of  $\epsilon$
\cite{Cakmur97}.  \figref{bistable-fig} shows two shadowgraph
images illustrating this bistability.

Experiments in circular cells with fluids of small Prandtl number
have suggested that above a particular value of $\epsilon$, SDC is the
state chosen for almost all initial conditions
\cite{deBruyn96,Liu96,Morris96}.  Just below this
value of $\epsilon$ a time-dependent, apparently chaotic state without
spirals was found.  Earlier experimental and theoretical
investigations of the latter state \cite{CrossHohenberg93}
led to the conclusion that the tendency of
rolls to align perpendicularly to the boundaries necessarily results in roll
curvature which, due to large-scale flows, leads to the observed
persistent pattern dynamics.  Motivated by these earlier observations,
investigations of SDC were mostly limited to 
circular containers.  Almost no experiments were conducted in
rectangular cells \cite{Morris96},  although pioneering experiments
by Gollub and coworkers  \cite{Gollub82}
had shown that ISR appeared after a long transient
in a small aspect ratio rectangular cell. 
In the remainder of this Letter, we present quantitative experimental
data from a large aspect ratio convection cell of square geometry
showing that the transition between SDC and ISR exhibits similarities
to equilibrium phase transitions in finite-size systems.

The experiment was conducted in a square cell with aspect
ratio $\Gamma = (L/2) / d = 50$, where $L$ is the length on a side.
The fluid was 
${\rm CO_2}$ gas at a pressure of $(41.593 \pm 0.007)$ bar
and Prandtl number $\sigma = \tau_T / \tau_\nu = 1.1$, where
$\tau_\nu$ is the vertical viscous timescale, 
$\tau_T = d^2/\kappa = 2.6\,{\rm s}$ is the vertical thermal diffusion
timescale, and $\kappa$ is the thermal diffusivity
\cite{topplate}.  The
experimental setup was similar to that described in
Ref.~\cite{deBruyn96}.  The cell's circular top and bottom
sapphire plates were $1\,{\rm cm}$ thick, were spaced
$(623 \pm 4)\,{\rm \mu m}$ apart, and were parallel to within 
$\pm 3\,{\rm \mu m}$  over the $10\,{\rm cm}$ diameter.
The top plate temperature was set at $(24.00 \pm 0.05){\rm^\circ C}$
by a circulating water bath and the bottom plate was heated
by an electric film heater.  Both temperatures were regulated
to $\pm 0.3\,{\rm mK}$, and the pressure was regulated to
$\pm 5 \times 10^{-3}\,{\rm bar}$.  For this experimental 
situation the parameters were sufficiently temperature
independent so the Boussinesq approximation could be applied
\cite{topplate,Boden91}.
The bottom sapphire plate was coated with aluminum to allow
the visualization of the pattern from above using the
shadowgraph technique \cite{deBruyn96}.  
Eight circular paper sheets were placed
between the top and bottom plates, and a square of size $100 d$ was cut
out of the center of the circular sheets to provide the
boundary of the convection cell \cite{boundaryexplanation}.  
The measured onset of
convection was $(2.03 \pm 0.02)\,{\rm K}$, which is in good agreement
with the theoretical prediction of $(2.04 \pm 0.02)\,{\rm K}$.

\begin{figure}   
\centerline{\epsfxsize=2.80in \epsfbox{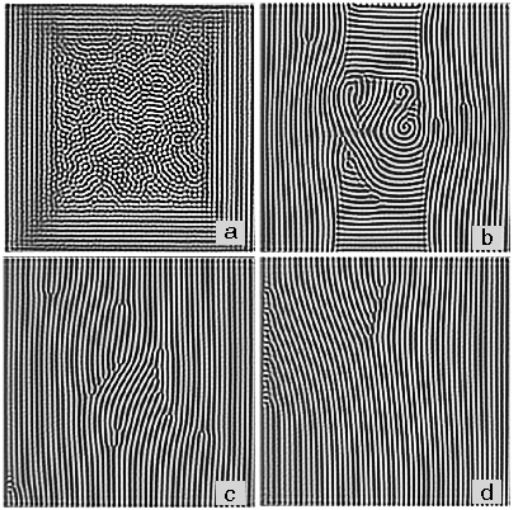}}
\bigskip
\caption{
Shadowgraph images of the convective pattern for a control parameter
jump from below onset ($\epsilon < 0$) to above onset ($\epsilon = 0.233$)
at times 
{\bf (a)} $\sim 45 \tau_T$,
{\bf (b)} $229 \tau_T$,
{\bf (c)} $458 \tau_T$, and
{\bf (d)} $22689 \tau_T = 9.1 \tau_h$, the stationary pattern.
}
\label{rollevolution-fig}
\end{figure}
\begin{figure}   
\centerline{\epsfxsize=2.85in \epsfbox{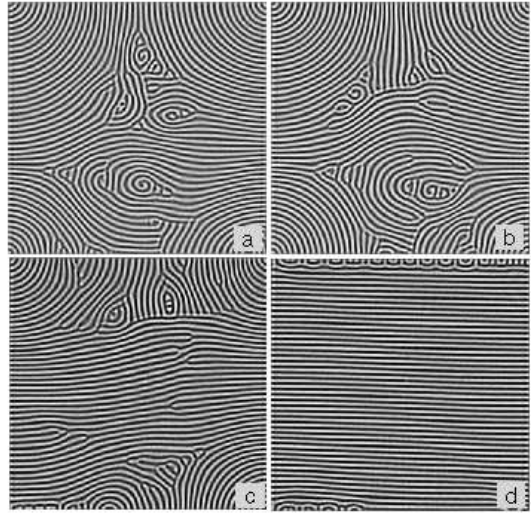}}
\bigskip
\caption{
Time evolution of the pattern
for $\epsilon = 0.58$ at times 
{\bf (a)} $13.70 \tau_h$,
{\bf (b)} $13.80 \tau_h$,
{\bf (c)} $14.07 \tau_h$, and
{\bf (d)} $28.96 \tau_h$, the stationary pattern.
}
\label{front-fig}
\end{figure}

For each value of the control parameter studied, the system
was initialized by a jump from below onset ($\epsilon < 0$)
to above onset ($\epsilon > 0$).  After this jump, straight rolls
initially formed near the sidewalls while a random pattern
appeared in the middle of the cell.  
For $\epsilon \lesssim 0.53$,
the pattern coarsened over time and developed after a 
transient into ISR with a few stray defects.  An example of
the pattern evolution  is
shown in \figref{rollevolution-fig}.  
 For $\epsilon \gtrsim 0.58$ 
the pattern did not order into ISR during
observation times of at least $20 t_h$, where 
$t_h = \Gamma^2 t_T$ is the horizontal thermal diffusion time.
In the intermediate regime,
$0.53 \lesssim \epsilon \lesssim 0.58$,
the behavior was more complicated with large patches of
almost ISR growing and competing with large patches of
SDC.  Sometimes, often after long transients, a patch of ISR grew
and filled the entire system.  As can be seen in \figref{front-fig},
once a patch of ISR connected two opposing sidewalls it 
grew via the propagation of  two almost flat
fronts. The behavior  described above  is reminiscent of that of
an equilibrium
finite-size system near a phase transition \cite{Newman93}.

We note that
the initial pattern evolution showed the brief appearance of spirals for
values of the control parameter as low as $\epsilon \approx 0.2$
(as shown in  \figref{rollevolution-fig}(b)).
This observation suggests that the bistability of
ISR and SDC may extend to lower control parameter values
than those reported here.  This conjecture would be in
agreement with earlier observations in a larger aspect ratio
circular cell ($\Gamma = 78$) in which SDC was found above
$\epsilon \approx 0.22$ for similar experimental conditions
\cite{Morris93}. It would also be consistent with 
both experiments \cite{Morris93,Liu96,Morris96,Hu95}
and simulations \cite{Decker94} in which it was found 
that the onset of SDC decreased with increasing system size.

\begin{figure}   
\centerline{\epsfxsize=2.25in \epsfbox{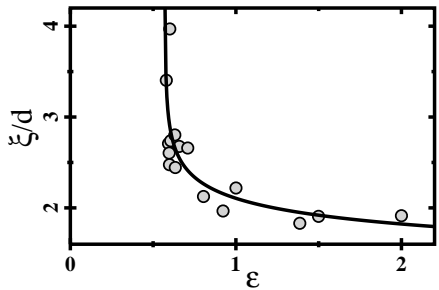}}
\bigskip
\caption{
Correlation length $\xi$ in the spatiotemporal chaotic regime.
}
\label{xi-fig}
\end{figure}

To quantify the transition from 
SDC to ISR, we  calculated several statistics in the vicinity of
the transition.  On the SDC side of the transition, we
measured the correlation length $\xi$ from the exponential
decay of the autocorrelation function:
\begin{equation}
C(\Delta \vec{x}) = 
\langle \left( u(\vec{x}+\Delta \vec{x}, t) - \langle u \rangle \right)
\left( u(\vec{x}, t) - \langle u \rangle \right) \rangle_{\vec{x},t},
\end{equation}
where $u(\vec{x}, t)$ is the intensity of the  
shadowgraph picture.  The autocorrelation function is
obtained from the Fourier transform of the power spectrum averaged
over the duration of each experimental run \cite{autocorr}.  As shown in
\figref{xi-fig}, the correlation length appears to diverge
at a {\sl finite} $\epsilon$.  
The solid line in \figref{xi-fig}
is given by $\xi = 1.9 d (\epsilon - 0.57)^{-0.12}$.  This result
is surprising since in an experiment with circular geometry
the correlation length was found to diverge as 
$\xi \propto \epsilon^{-0.43}$ \cite{Morris93,Morris96}. 
It should be noted, however, that in the same circular
experiment \cite{Morris96}
the measured data for the correlation time  was consistent
with a divergence at finite $\epsilon$, with an apparent
transition between two chaotic states.
In our experiment, however,  we observed a transition between
a spatiotemporal chaotic state and a stationary state. 
These apparent discrepancies
may be explained by the incompatibility of
the circular geometry with ISR, as discussed above.

To further quantify the transition between SDC and ISR,
we have calculated the spectral
pattern entropy \cite{Neufeld94,Xi94}.  This quantity is
defined through a spectral distribution function:
\begin{equation}
p(\vec{k}, t) = \frac{|\Psi(\vec{k}, t)|^2}{\int d^2k |\Psi(\vec{k}, t)|^2},
\end{equation}
where $\Psi(\vec{k}, t)$ is the Fourier transform of the two dimensional
pattern at time $t$.  Using this distribution function, the
spectral pattern entropy $S(t)$ can be defined as:
\begin{equation}
S(t) = - \int d^2k\, p(\vec{k}, t) \ln p(\vec{k}, t).
\end{equation}
$S(t)$ measures the disorder in the pattern.  For example, if the pattern
is ideal (only one mode is excited) $S = 0$ and otherwise $S > 0$.

\begin{figure}   
\centerline{\epsfxsize=2.85in \epsfbox{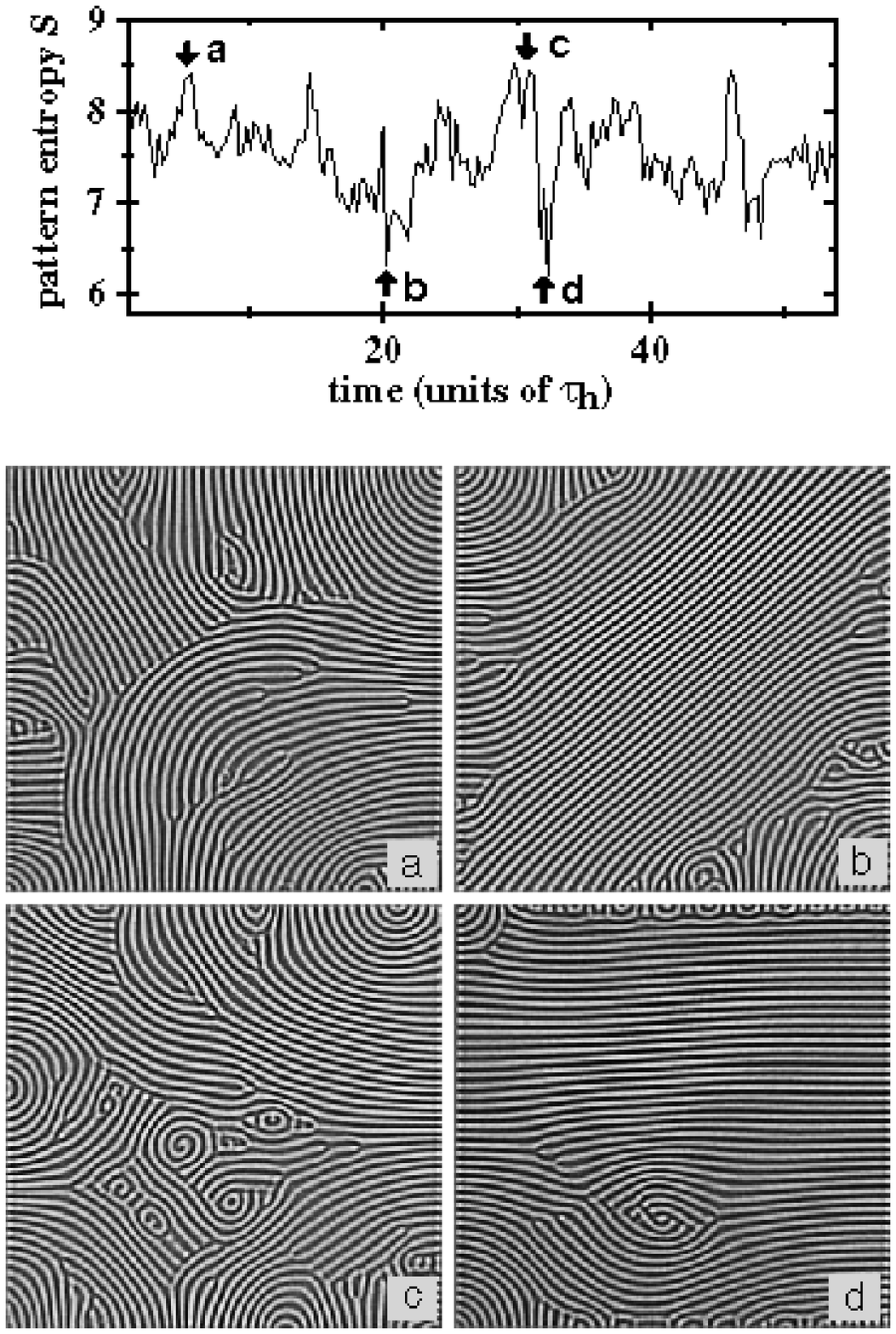}}
\bigskip
\caption{
Time evolution of the pattern entropy $S(t)$ and the pattern
for $\epsilon = 0.554$ at times
{\bf (a)} $14.52 \tau_h$,
{\bf (b)} $20.27 \tau_h$,
{\bf (c)} $30.95 \tau_h$, and
{\bf (d)} $32.32 \tau_h$.
}
\label{entropytime-fig}
\end{figure}

In \figref{entropytime-fig} the time evolution of the pattern entropy
is shown for $\epsilon = 0.554$.
The patterns at the
marked places show a clear correlation with the value of the pattern
entropy.  As can be seen in Figs.~\ref{entropytime-fig}(a) and
\ref{entropytime-fig}(c), patterns with disordered or curved regions
lead to large values of $S(t)$, while patterns with straight
regions, such as those in Figs.~\ref{entropytime-fig}(b) and
\ref{entropytime-fig}(d), lead to small values of $S(t)$.  The latter
two patterns also show that during the
evolution, well-ordered regions were typically aligned 
either diagonally or perpendicularly to one of the sidewalls of the cell;
it appeared that the pattern was probing the system's symmetries.

\begin{figure}   
\centerline{\epsfxsize=2.25in \epsfbox{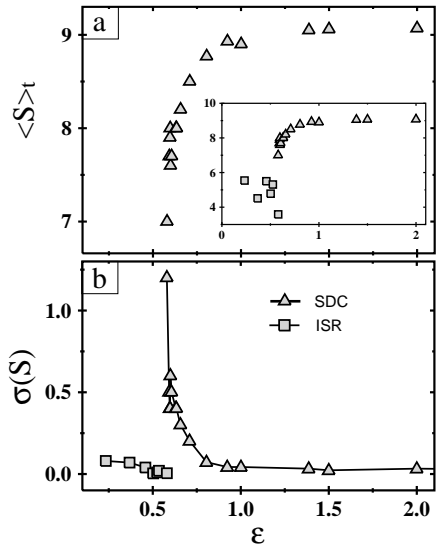}}
\bigskip
\caption{
{\bf (a)} The time-averaged pattern entropy $\langle S \rangle_t$
and {\bf (b)} variance $\sigma(S)$ as a function of $\epsilon$.
The inset in {\bf (a)} shows data for both SDC and  ISR.
}
\label{entropystats-fig}
\end{figure}

The experimental data are summarized in \figref{entropystats-fig} \cite{note}.
\figref{entropystats-fig}(a) shows the temporal average of the pattern entropy
$\langle S \rangle_t$ as a function of $\epsilon$.  
As the transition to ISR is approached from above,
$\langle S \rangle_t$ shows a sharp decrease.
\figref{entropystats-fig}(b) shows the variance $\sigma(S)$
over the same range of $\epsilon$.  As the transition is
approached, $\sigma(S)$ displays a sharp increase.
The location of the sharp changes is consistent with the
divergence of the correlation length in \figref{xi-fig}. For larger $\epsilon$ 
the variance  $\sigma(S)$ approaches a small
value,  suggesting that the system consists of many independent,
fluctuating  subsystems. 
Again, the behavior shown in \figref{entropystats-fig} appears to be
similar to that of equilibrium phase transitions \cite{Newman93}.

For a fluid of Prandtl number $\sigma = 1.1$ in a square convection
cell of large aspect ratio,
we have found a competition between spiral defect chaos
and ideal straight roll states.
This competition is particularly evident in the interplay
of disordered regions and straight roll regions during the evolution
of the SDC states.
Using spatial correlation lengths and
statistics based on the spectral pattern entropy, we have provided
quantitative evidence for a transition between SDC and ISR at a
finite value of $\epsilon$.  This transition shows many similarities
to transitions in finite-size equilibrium systems, including an
intriguing intermittent behavior near the transition.
We are currently
investigating whether experiments in cells of different aspect ratios
can be used to predict the onset of SDC in infinitely extended systems.

We thank W.~Pesch and J.~Sethna for many fruitful
discussions. We are grateful to R.~Ragnarsson who
provided essential programming for the operation
of the experiment. This work was
supported by the National Science
Foundation through Grants DMR-9320124 and ASC-9503963.
E.B. acknowledges support from the Alfred P. Sloan Foundation,
D.A.E. from the Cornell Theory Center, and B.B.P. from the
Department of Education.


\begin{thebibliography}{10}
\bibitem{topplate} Since the top plate temperature was fixed and
the bottom temperature was varied, the mean temperature and thus the 
fluid parameters changed for each $\epsilon$.  For each mean
temperature a material parameter program \cite{deBruyn96} was used to
calculate the critical
temperature difference and the other parameters.
We found  $\sigma = 1.103 - 0.016 \epsilon$,
$\tau_T = 2.630 - 0.071 \epsilon$, and the non-Boussinesq parameter
${\cal P} = - 0.481 + 0.024 \epsilon$.
\bibitem{boundaryexplanation} At two opposing sides of the square,
the paper was cut in such a way that the top four sheets were recessed
by 1 mm, while the other sides were not modified.  We found no
experimental evidence that the unequal boundaries caused the ISR 
state to be preferentially aligned. In the figures the stepped
boundaries are at the left and right and the non-stepped
are at the top and bottom.
\bibitem{autocorr} In the vicinity of the transition from SDC to ISR,
the autocorrelation  function was spatially anisotropic.  We report lengths
computed in the direction in which the correlations decayed slowest.
\bibitem{note}
In \figref{entropystats-fig}, the points labeled as SDC only include 
those experimental runs for which SDC appeared to be the 
asymptotic state.  It is interesting to note that in the intermittent
regime, $0.53 \lesssim \epsilon \lesssim 0.58$, statistics
measured during the {\sl disordered} portion of the transient
are consistent with the plotted data.
\end{thebibliography}
\end{document}